\newcommand{\f}[1]{Fig.~\ref{#1}}

\def\be{\begin{equation}}
\def\ee{\end{equation}}
\def\bea{\begin{eqnarray}}
\def\eea{\end{eqnarray}}
\def\l({\left(}
\def\r){\right)}

\def\s2{w/\sqrt{2}}

\documentstyle[prb,aps,multicol,graphicx,hhline,psfig]{revtex}
  \renewcommand{\narrowtext}{\begin{multicols}{2} \global\columnwidth20.5pc}
  \renewcommand{\widetext}{\end{multicols} \global\columnwidth42.5pc}

\begin{document}
\title{Current-induced dendritic magnetic instability in superconducting MgB$_2$ films}
\author{A. V. Bobyl$^{1,2}$, D.~V. Shantsev$^{1,2}$,
T.~H.~Johansen$^{1,}$\cite{0}, W. N. Kang$^3$, H. J. Kim$^3$, E. M. Choi$^3$, 
S. I. Lee$^{3,}$\cite{1}
}
\address{
$^1$Department of Physics, University of Oslo, P. O. Box 1048
Blindern, 0316 Oslo, Norway\\
$^2$A. F. Ioffe Physico-Technical Institute, Polytekhnicheskaya 26,
St.Petersburg 194021, Russia\\
$^3$National Creative Research Initiative Center for Superconductivity, 
Department of Physics, Pohang University of 
Science and Technology, Pohang 790-784, Republic of Korea
}
%\date{\today}
\maketitle

\vspace{-4.2cm}
\begin{center}
{
Resubmitted to Appl. Phys. Lett. on 15.04.2002%, cond-mat/0201260
}
\end{center}
\vspace{3.2cm}

\begin{abstract}
Magneto-optical imaging reveals that in superconducting films of MgB$_2$ a 
pulse of %added 
transport current
creates avalanche-like flux dynamics where highly branching dendritic
%penetration 
patterns are formed. The instability is triggered when the 
current exceeds a threshold value, and the superconductor, shaped as a long strip, 
is initially in the critical state.
The instability exists up to 19~K, 
which is a much wider temperature range than in previous experiments, where 
dendrites were formed by 
%applying 
a
slowly varying %added 
magnetic field. 
The instability is believed to be of thermo-magnetic origin indicating that 
thermal stabilization may become crucial in applications of MgB$_2$.
\end{abstract}

%\pacs{PACS numbers: 74.76.-w, 74.25.Ha, 74.60.Ge, 68.60.Dv} 
%74.76.Bz, 74.60.Jg 

\narrowtext

%\subsubsection{Introduction}

\vspace{1cm}

Following the discovery\cite{r1} of superconductivity below 39~K in 
polycrystalline MgB$_2$ a tremendous effort is now being invested to produce
high-quality films for basic studies and for potential industrial use.
Of primary importance for technological applications is the magnitude and
stability of the critical current density $J_c$, or more generally, the
static and dynamical behavior of the magnetic vortices.
Whereas $J_c$ as high as 10$^7$~A/cm$^2$ have already been reported
for MgB$_2$ thin films\cite{r2,prl} it has also been shown that this 
promising value and the technological potential may be 
challenged by a surprising complexity in the flux dynamics. 
Magneto-optical (MO) imaging recently revealed\cite{prl,sust} that
below 10 K the flux penetration in MgB$_2$ films is dominated by large 
dendritic structures abruptly entering the film, thus dramatically 
contrasting the smooth and regular flux dynamics normally found in 
type-II superconductors. 
This observation is consistent with the large fluctuations reported in the 
magnetization for various MgB$_2$ samples.\cite{r3,r4,r5,mumtaz} 
The abrupt drops appearing in the magnetization curve, 
much like at flux jumps\cite{wipf-prb,r6} %added 
only in very large numbers, 
give an effective 50\% suppression of the apparent $J_c$.\cite{prl}

Abrupt flux dynamics with dendritic penetration patterns has been observed
earlier also in films of Nb\cite{r10,r11,r12} and YBCO.\cite{r13,r14}
In all the previous studies the dendrites or magnetization noise have been 
the result of a gradual increase or decrease of the applied magnetic field 
(or a short laser pulse perturbing the remanent state\cite{r13,r14}). It is therefore still an open question 
-- one of vital importance for many practical applications -- whether flux 
dendrites will be formed also in response to a transport current. 
In this work it is shown that in MgB$_2$ thin films, a pulse of transport 
current indeed can trigger the dendritic instability, and we use MO imaging
to reveal details of this behavior. 
 
%\subsubsection{Sample}
 
A film of MgB$_2$ was fabricated on ($1\bar{1}02$) Al$_2$O$_3$ substrate using 
pulsed laser deposition. An amorphous B film was first deposited, and then 
sintered at high temperature in a Mg atmosphere.\cite{r2} 
The film had transition at  
39~K with width 0.7~K, and a high degree 
of $c$-axis alignment perpendicular to 
the film plane. The film thickness was 300~nm, and its
lateral dimensions were 3$\times$10~mm$^2$. The strip was at both ends
equipped with a contact pad of size 3$\times$1.5~mm$^2$ fabricated 
by sputtering of Au through a mask.
Copper wires were attached to the pads using silver paste. 
%R1208 Bio-Rad, giving a contact resistance less than 0.02~$\Omega$. 
The MO imaging was performed 
with the sample mounted on the cold finger in an Oxford Microstat--He optical 
cryostat. The MO sensing element, a plate of in-plane magnetization 
ferrite garnet film, was placed on the MgB$_2$ film covering the area between the contacts.
%The resistance of external wires and vacuum switches was less than 0.12~$\Omega$.
The pulses of transport current 
with rise time of 0.1~$\mu$s were sent through the MgB$_2$ film
using 
a power transistor
%To create pulses of transport current a power transistor
%(JPG4-50)
controlled by a
pulse generator. 
%The current pulses sent through the MgB$_2$ film had a
%rise time of 0.1~$\mu$s, and a duration adjustable from 1.5 to 15~$\mu$s.
%Overheating of the sample was avoided by using a D518E stabilitron (13V).
The MO images were recorded %10~ms 
immediately after the current pulse, the camera exposure time being 250~ms. %added
%, using syncronized triggering from an Advantech PCL 818L card.
The time development of the $I-V$ characteristics %of the sample
during the pulse 
measured by a two-point scheme
was monitored via a TDS-420 oscilloscope.

%\subsubsection{Results and Discussion}

First, MO images were taken of the MgB$_2$ film after zero-field-cooling (ZFC)
to 3.5~K and applying a perpendicular magnetic field. Results of this 
reference experiment (with no transport current) are shown in \f{f_f1}.
In the images both for increasing (a) and decreasing field (b)
one clearly sees that the flux penetration picture is dominated by dendritic
patterns. The dendrites are formed instantaneously in response to a slowly varying
applied field. Like in the thicker MgB$_2$ film studied in Ref.~\onlinecite{prl},
also this sample has a threshold temperature of 10~K, above which
the flux penetration proceeds in a conventional way with a gradual advancement of
flux front.

Shown in \f{f_f2}(a,b) are flux distributions recorded at 12.5~K, 
as the applied field cycled from 0 to 70~mT, and back to 
the remanent state. Evidently, no dendrites were formed, and the 
final state image, (b), shows all the characteristics of a critical 
remanent state, where the field is maximum at the so-called 
discontinuity lines.\cite{dlines} % added ref
They are seen as the most bright lines in the image, one in 
the center of the strip and two pointing towards the corners. 
At these lines the current, which flows in rectangular loops, 
as indicated by the white arrows, changes its 
direction discontinuously.
% added:
Even without dendrites the flux distribution is not smooth, especially
near the edges, see also \f{f_f1}(a), indicating non-uniformity of the film. 
 
%Fig.1
\begin{figure}
\centerline{\includegraphics[width=8.5cm]{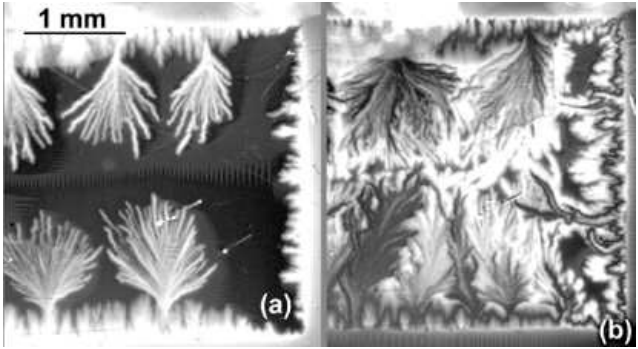}}
%\centerline{\psfig{figure=ha.eps,width=8.5cm}}
\vspace{0.2cm}
\caption{Magneto-optical images of the flux distribution in a rectangular strip of MgB$_2$ thin film at $3.5$~K. The image brightness represents the magnitude of the
local flux density.
(a) Several branching dendritic flux structures have been created by
a 10~mT field applied to the ZFC film.
(b) Remanent state after a maximum applied field of 70~mT.
Dark dendrites containing antiflux are formed on top of the bright dendrites
created while the field was increased.
Both images were taken before current contacts were attached to the film.\label{f_f1}}
\end{figure}

With the sample prepared in the remanent state, (b), a pulse of transport current was applied. 
As a result, dendritic flux structures of the type seen in (c) and (d) were formed. 
In (c) and (d) the current pulse was passed through the strip in opposite
directions, as indicated by black arrows. Interestingly, one observes that the
dendrites always develop from the side of the film where the transport current adds
up with the remanent state persistent current. Furthermore, 
%the images show that 
the %current-induced 
dendrites tend to expand to the middle of the film, in agreement with
numerical simulations.\cite{aranson} 
%into the regions with highest flux density. 
This contrasts the virgin penetration seen in \f{f_f1}(a), where the %field-induced 
dendrites develop more freely into the Meissner state region.

Similar experiments performed at different temperatures showed that a 
%transport
current 
pulse %added
will trigger flux dendrites up to 19~K, but never above.
This threshold temperature is approximately twice that of dendrite
formation induced by a 
slowly changing ($\sim$ 0.01~T/s) %added 
applied field. 
 %added :
The difference in threshold temperature can be qualitatively explained by
the sensitivity of thermo-magnetic instability to the field (or current)
ramp rate that determines magnitude of the electric field in the sample.\cite{r6,wipf-prb}  
The dendrite pattern was insensitive to a change in
the pulse duration 
in the range 1.5 -- 15~$\mu$s
if the rise time was kept constant.
This also suggests that all dendrites are formed during the rise time.

%Fig.2

In another series of experiments the height of the current pulse, $I$, was varied.
By applying the pulse to the same remanent state, \f{f_f2}(b), we found that
dendrites occur only when $I$ exceeds a certain magnitude. In the present case, 
the threshold current equals 7.0~A at $T=12.5$~K, and depends only weakly on temperature.
We also observe that the more $I$ exceeds the threshold, the larger 
the dendritic structure becomes in size.
Eventually, for sufficiently large $I$ two dendrites are formed during the pulse.
The full scenario is illustrated in \f{f_f3},%
\begin{figure}
\centerline{\includegraphics[width=6.28cm]{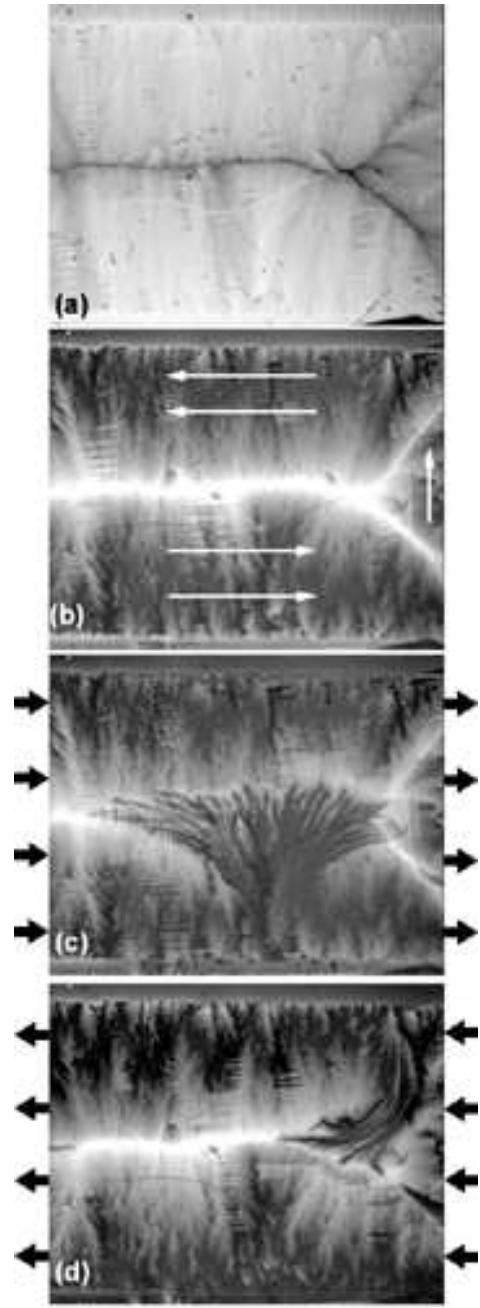}}
%\centerline{\psfig{figure=f2s.eps,width=6.8cm}}
%\vspace{2cm}
\caption{MO images of the MgB$_2$ film at $12.5$~K.
(a) Applying a field of 34~mT after ZFC.
(b) Remanent state after having raised the field to a maximum of 70~mT.
White arrows indicate the direction of the shielding currents in the
rectangular film.
(c,d) After applying a 8.5~A current pulse (duration 3~$\mu$s) to the remanent state,
(b), with opposite current directions, see black arrows.
The pulse creates dendritic flux structures, which always develop
from the edge where the transport and shielding currents add up constructively.
% new:
The pulse also leads to marginal flux exit seen as slight darkening 
of the opposite edge; this effect is discussed in more detail
in other 
theoretical\cite{zeldov} and experimental\cite{strip-cc} works.  
Zigzag lines are caused by domain walls in the MO film and should
be ignored.   
\label{f_f2}}
\end{figure}
\noindent which shows by solid circles
the area covered by the dendrite(s). The vertical arrows indicate at which
$I$ the number of dendrites is incremented from zero (below 7~A) to three (at 17~A).
For very large currents heating near 
the contact pads affects the results. 
It was detected for $I > 15$~A by the appearance of 
a non-stationary voltage across the sample 
which increased %added
during the pulse, 
see \f{f_f3}. In addition, flux-free areas around the contacts 
were seen in the MO images.

Current-induced dendrites were formed not only by starting from the
critical remanent  state, but also from critical states like the one in \f{f_f2}(a).
The dendrites develop then from the opposite side of the strip since
the flow of screening currents is here reversed compared to the
remanent state. Hence, all these results show consistently that
the dendritic instability is triggered when the transport and the
critical-state shielding current flow in the same direction.
All attempts to create dendrites by passing a current pulse through
a virgin ZFC film failed for $I$ up to 30 A, even at the lowest
temperature of 3.5~K.

It is believed that the dendritic avalanche behavior results
from a thermo-magnetic instability in the superconductor. 
Vortex dynamics simulations\cite{prl,aranson} and experiments on MgB$_2$ films under
different thermal conditions\cite{phc} suggest that
the instability stems from the local heating produced by flux motion.
The heating will facilitate flux motion nearby, which in turn can
lead to a large-scale avalanche invasion of depinned flux lines.
The same mechanism is responsible for flux jumps found in the Tesla-range
of the magnetization loop of bulk materials.\cite{r6}
The present investigation is the first one
to demonstrate that the dendritic type of
the thermo-magnetic instability
can be triggered by a transport current.
Moreover, films of MgB$_2$ are found
more susceptible%
\begin{figure}
\centerline{\includegraphics[width=8.5cm]{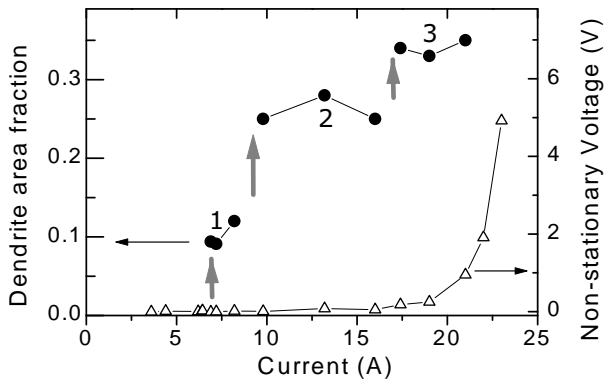}}
%\centerline{\psfig{figure=f3.eps,width=8.5cm}}
\caption{The area covered by the current-induced dendrites (solid circles) at $T=12.5$~K. 
The vertical arrows indicate at which currents the number of dendrites increases by one.
Open triangles show 
%a difference in the voltage measured 1.0 and 2.7~$\mu$s after the pulse start
an increase in the voltage from 1.0 to 2.7~$\mu$s after the pulse start 
(the pulse duration was 3~$\mu$s).  
\label{f_f3}}
\end{figure}
\noindent for the current-induced instability
than for instability induced by varying the applied field.
This shows that careful design of thermal 
stabilization may become essential in order to make viable
devices out of MgB$_2$ films.

The work was financially supported by The Norwegian
Research Council, and by the Ministry of Science and Technology of
Korea through the Creative Research Initiative Program.

%\newpage
%\newpage
%
%\widetext
%\narrowtext
%
%\newpage

\widetext

\vspace{-0.1cm}
\begin{references}
\vspace{-1cm}

\bibitem[*]{0}Email for correspondence: t.h.johansen@fys.uio.no

\bibitem[\dag]{1}Email for request of the materials: silee@postech.ac.kr

\bibitem{r1} J. Nagamatsu, N. Nakagawa, T. Muranaka, Y. Zenitani, J.
Akimitsu,.  Nature {\bf 410}, 63 (2001).
%\bibitem{r2} W. N. Kang, H. J. Kim, E. M. Choi, C. U. Jung, S. I. Lee,
%Science  {\bf 292}, 1521 (2001). 10.1126/science.1060822.
\bibitem{r2} H. J. Kim, W. N. Kang, E. M. Choi, M. S. Kim, K. H. P. Kim, S. I. Lee,
Phys. Rev. Lett.  {\bf 87}, 087002 (2001).
\bibitem{prl} T.H. Johansen, M. Baziljevich, D.V. Shantsev, P.E. Goa, Y.M. Galperin, W.N. Kang, H.J.
    Kim, E.M. Choi, M.-S. Kim, S.I. Lee, cond-mat/0104113.
%submitted to Europhys. Lett.
\bibitem{sust} T.H. Johansen, M. Baziljevich, D.V. Shantsev, P.E. Goa, 
Y.M. Galperin, W.N. Kang, H.J. Kim, E.M. Choi, M.-S. Kim, S.I. Lee, 
Supercond. Sci. Technol. {\bf 14}, 726-728 (2001). 
\bibitem{r3} S. Jin, H. Mavoori, C. Bower, R. B. van Dover, Nature {\bf
411}, 563  (2001).
\bibitem{r4} Z. W. Zhao, S. L. Li, Y. M. Ni, H. P. Yang, Z. Y. Liu, H. H. Wen,
W. N. Kang, H. J. Kim, E. M. Choi, and S. I. Lee, 
Phys. Rev. B {\bf 65}, 064512 (2002).
\bibitem{r5} S. X Dou et al., Physica C {\bf 361}, 79 (2001).
\bibitem{mumtaz} A. Mumtaz, W. Setyawan, and S. A. Shaheen, Phys. Rev. B {\bf 65}, 020503(R) (2002).
%added
\bibitem{wipf-prb} S. L. Wipf, Phys. Rev. {\bf 161}, 404 (1967).
\bibitem{r6} R.G. Mints, A.L. Rakhmanov, Rev. Mod. Phys. {\bf 53}, 551-
592 (1981).

\bibitem{r10} M. R. Wertheimer, J de G. Gilchrist, J. Phys. Chem Solids
{\bf 28},  2509 (1967).
\bibitem{r11} C. A. Duran, P. L. Gammel, R. E. Miller, D. J. Bishop,
Phys. Rev. B  {\bf 52}, 75 (1995).
\bibitem{r12} V. Vlasko-Vlasov, U. Welp, V Metlushko, G. W. Crabtree,
Physica C  {\bf 341-348}, 1281 (2000).
\bibitem{r13} P. Leiderer, J. Boneberg, P. Bruell, V. Bujok, S.
Herminghaus, Phys.  Rev. Lett. {\bf 71}, 2646 (1993).
\bibitem{r14} U. Bolz, J. Eisenmenger, J. Schiessling, B.-U. Runge, P.
Leiderer,  Physica B {\bf 284-288}, 757 (2000).
\bibitem{zeldov} E. Zeldov,  J. R. Clem, M. McElfresh,
and M. Darwin,  Phys. Rev. B {\bf 49}, 9802 (1994).
\bibitem{strip-cc} A. V. Bobyl, D. V. Shantsev, Y. M. Galperin, T. H. Johansen, 
M. Baziljevich, S. F. Karmanenko, 
Supercond. Sci. Technol. {\bf 15}, 82 (2002). 
\bibitem{dlines} Th. Schuster, M. V. Indenbom, M. R. Koblischka, 
H. Kuhn, and H. Kronm\"{u}ller, Phys. Rev. B {\bf 49}, 3443 (1994).  
\bibitem{aranson} I. Aranson, A. Gurevich, V. Vinokur, Phys. Rev. Lett.
{\bf 87}, 067003 (2001).
\bibitem{phc} M. Baziljevich, A. V. Bobyl, D.V. Shantsev, E. Altshuler,
T.H. Johansen and S.I. Lee,
Physica C %added 
{\bf 369}, 93 (2002). 

\end{references}
\end{document}